\documentstyle[preprint,aps]{revtex}
\begin{document}
\draft
\preprint{IP/BBSR/94-15}
\title{VACUUM STRUCTURE IN QCD WITH QUARK AND GLUON CONDENSATES}
\author{A. Mishra, H. Mishra, Varun Sheel}
\address{Theory Group, Physical Research Laboratory, Navrangpura, Ahmedabad
380 009, India}
\author{S.P. Misra and P.K. Panda}
\address{Institute of Physics, Bhubaneswar-751005, India.}
\maketitle
\begin{abstract}
We consider here the vacuum structure in QCD with both quark
and gluon condensates and a variational ansatz
for the ground state.
The method is nonperturbative using only
equal time algebra for the field operators.
We then find that a constrained energy minimisation of the Hamiltonian
leads to a QCD vacuum with both quark and gluon condensates
for $\alpha_s > \alpha_c=0.62$.
Pion decay constant and the charge radius of the pion seem to fix the QCD
coupling constant $\alpha_s$ as 1.28. The approach opens up possibilities
of relating the mysterious vacuum structure with common place
hadronic properties.
\end{abstract}
\narrowtext
\section{Introduction}
It is now believed that quantum chromodynamics (QCD) is the correct
theory for strong interaction physics of quarks and gluons.
At low energies, however, the coupling constant in QCD
becomes large leading to breakdown of the perturbative
calculations. In this regime, the vacuum structure is also known to be
nontrivial \cite{suryak} with nonzero expectation values for
quark and gluon condensates \cite{svz}.
Instability of QCD vacuum with constant chromomagnetic
field or with vortex condensate formation has been studied since quite
some time with a semiclassical approach \cite{nielsen}.
QCD vacuum has also been studied with gluon
or  glueball condensates \cite{schutte,hans}
as well as with nonperturbative solutions to Schwinger Dyson
equations \cite{corn}. Further, a nontrivial vacuum structure with quark
condensates in Nambu Jona Lasinio type of models \cite{nambu}
has been seen to be consistent with
low energy hadron physics.
It is therefore desirable to examine the vacuum stucture in QCD
with  {\em both} quark and gluon condensates.

We had proposed earlier a variational method
which is nonperturbative with an explicit structure for the QCD
vacuum. This has been
applied to the case of gluon condensates for vacuum structure
in $SU(3)$ Yang-Mills fields demonstrating the instability of
perturbative vacuum when coupling is greater than a
critical value \cite{pramana}.
Same methods have been applied to study the vacuum structure with
quark condensates in QCD motivated phenomenological effective potential
models \cite{pot} and in Nambu-Jona-Lasinio model \cite{njl}. For the
ground state or vacuum this has been achieved through a minimisation of energy
density, free energy density or thermodynamic potential depending on
the physical situation.
For the chiral symmetry breaking
we had also considered a simple ansatz of taking the perturbative
quarks having a phenomenological gaussian distribution in the
nonperturbative vacuum \cite{amspm}.
This appeared to describe a host of low energy hadronic properties
as being related to the vacuum structure associated with chiral
symmetry breaking.
However, here no energy minimisation has been attempted.
In the present paper we shall analyse the vacuum structure in QCD
with both quark and gluon condensates and discuss its stability as
opposed to taking {\em only} gluon condensates
\cite{svz,nielsen,schutte,hans,corn,pramana}
or {\em only} quark condensates \cite{nambu,pot,njl}.

As earlier, we shall take specific forms of condensate functions
for quarks and gluons
to describe a trial state for destabilised vacuum.
Such an  ansatz necessarily has
limited dynamics. We shall  circumvent this partially
with the constraints that the
SVZ parameter $\frac{\alpha_s}{\pi}<G^a_{\mu\nu}{G^a}^{\mu\nu}>$ and
the pion decay constant $c_\pi$ shall be correctly reproduced as experimentally
known. The condensate functions still contain two parameters, over which energy
is minimised. We then note that for $\alpha_s$ greater than a
critical coupling $\alpha_c$, perturbative vacuum destabilises with
nonvanishing condensates in {\em both} quark and gluon sectors.

We organise the paper as follows. In section II, we briefly
recapitulate quantisation of QCD in Coulomb gauge and give an
explicit construct for the nonperbutative vacuum with
quark and gluon condensates.
The ansatz for the QCD vacuum is similar to BCS
ansatz of Cooper pairs in the context of superconductivity.
In section III we consider the stability of such a trial state
through an energy minimisation where pion decay constant
and SVZ parameter are taken as constraints as above and discuss the results.
In section IV we summarise our conclusions. We also notice that
for $\alpha_{s} \simeq 1.28$, the ansatz function of Ref. \cite{amspm}
along with the correct charge radius of the pion
is reproduced.

The method considered here is  nonperturbative as we shall be
using the equal time quantum algebra for the field operators, but
is limited by the choice of the ansatz functions. The technique has been
applied earlier to solvable cases \cite{hmgrnv} to examine its reliability
and to ground state structure
for electroweak symmetry breaking and cosmic rays
\cite{higgs} and to nuclear matter, deuteron or quark stars \cite{nmtr}.

\section {Vacuum in QCD with quark and gluon condensates}
The QCD Lagrangian is given as
\begin{equation}
{\cal L} ={\cal L}_{gauge}+{\cal L}_{matter}+{\cal L}_{int},
\end{equation}
\noindent where
\begin{mathletters}
\begin{equation}
{\cal L}_{gauge}=-{1\over 2}G^{a\mu\nu}(\partial_{\mu}{W^{a}}_{\nu}
-\partial_{\nu}{W^{a}}_{\mu}+gf^{abc}{W^{b}}_{\mu}{W^{c}_{\nu})}
+{1\over 4}{G^{a}}_{\mu\nu}{G^{a\mu\nu}},
\end{equation}
\begin{equation}
{\cal L}_{matter}=\bar \psi (i \gamma ^\mu \partial _\mu)\psi
\end{equation}
\noindent and
\begin{equation}
{\cal L}_{int}=g\bar \psi \gamma ^\mu  \frac {\lambda^a}{2}
W _\mu^a\psi,
\end{equation}
\end{mathletters}
\noindent where ${W^{a}}_{\mu}$ are the SU(3) colour gauge fields.
We shall quantise in Coulomb gauge \cite{schwinger}
and write the
electric field  ${G^{a}}_{0i}$ in terms of
the transverse and longitudinal parts as
\begin{equation}
{G^a}_{0i}=
{^TG^a}_{0i}+{\partial_{i}{f}^{a}},
\end{equation}
where ${f}^{a}$ is to be determined.
We take at time t=0  \cite{pramana}
\begin{mathletters}
\begin{equation}
{W^{a}}_{i}(\vec x)={(2\pi)^{-3/2}}\int
{d\vec k\over \sqrt{2\omega(\vec k)}}({a^a}_{i}(\vec k) +
{{a^a}_{i}(-\vec k)}^{\dagger})\exp({i\vec k.\vec x})
\end{equation}
 \noindent and
 \begin{equation}
 {^{T}G^{a}}_{0i}(\vec x)={(2\pi)^{-3/2}} i \int
{d\vec k}{\sqrt{\omega(\vec k)\over 2}}(-{a^a}_{i}(\vec k) +
{{a^a}_{i}(-\vec k)}^{\dagger})\exp({i\vec k.\vec x}),
\end{equation}
\end{mathletters}
where, $\omega(k)$ is arbitrary \cite{schwinger}
and for equal time algebra we have
\begin{equation}
\left[ {a^a}_{i}(\vec k),{{a^b}_{j}(\vec k^{'})}^\dagger\right]=
\delta^{ab}\Delta_{ij}(\vec k)\delta({\vec k}-{\vec k^{'}}),
\end{equation}
with
 \begin{equation}
 \Delta_{ij}(\vec k)={\delta_{ij}}-{k_{i}k_{j}\over k^2}.
\end{equation}
The equal time quantization condition for the fermionic sector
is given as
\begin{equation}
[\psi _{\alpha}^i(\vec x,t),\psi _\beta^j (\vec y,t
)^{\dagger}]_{+}=\delta ^{ij}\delta _{\alpha \beta}\delta(\vec x -\vec y),
\end{equation}
where $i$ and $j$ refer to the colour and flavour indices
\cite{pramana,pot}.
We now also have the field expansion for fermion field
$\psi$ at time t=0  given as \cite{pramana,pot}
\begin{equation}
\psi^{i}(\vec x)=\frac {1}{(2\pi)^{3/2}}\int
\left[U_r(\vec k)c^{i}_{Ir}(\vec k)
+V_s(-\vec k)\tilde c^{i}_{Is}(-\vec k)\right]
e^{i\vec k\cdot \vec x} d\vec k,
\end{equation}
where $U$ and $V$ are given by \cite{spm78}
\begin{equation}
U_r(\vec k)=\frac{1}{\sqrt 2}\left( \begin{array}{c}1
\\ \vec \sigma \cdot\hat k
\end{array}\right)u_{Ir} ;\quad V_s(-\vec k)=\frac{1}{\sqrt 2}\left(
\begin{array}{c}\vec \sigma .\hat k
\\ 1\end{array}\right)v_{Is},
\end{equation}
\noindent for free chiral fields.
The above are consistent with the equal time anticommutation conditions
with \cite{spm78}
\begin{equation}
[c^{i}_{Ir}(\vec k),c^{j}_{Is}(\vec k')^\dagger]_{+}=
\delta _{rs}\delta^{ij}\delta(\vec k-\vec k')=
[\tilde c_{Ir}^{i}(\vec k),\tilde c_{Is}^{j}(\vec k')
^\dagger]_{+},
\end{equation}

In Coulomb gauge, the expression for the Hamiltonian
density, ${\cal T}^{00}$ from equation (1) is given as \cite{schwinger}
\begin{eqnarray}
{\cal T}^{00}&=&:{1\over 2}{^{T}{G^a}_{0i}}{^{T}{G^a}_{0i}}+
{1\over 2}{W^a}_{i}(-\vec \bigtriangledown^2){W^a}_{i}+
gf^{abc}W^a_iW^b_j\partial _i W^c_j\nonumber \\ &+&
{{g^2}\over 4}f^{abc}f^{aef}{W^b_i}{W^c_j}{W^e_i}{W^f_j}+
{1\over 2}(\partial_{i}f^{a})(\partial_{i}f^{a})
\nonumber \\ &+&\bar \psi (-i \gamma ^i \partial _i)\psi
-g\bar \psi \gamma ^ i \frac {\lambda^a}{2}
W _i^a\psi:,
\label{t00}
\end{eqnarray}
\noindent where : : denotes the normal ordering with respect to
the perturbative vacuum, say $\mid vac>$, defined through
${a^a}_{i}(\vec k)\mid vac>=0$, $c_{Ir}^i(\vec k)\mid vac>=0$
and $\tilde c_{Ir}^i(\vec k)^\dagger\mid vac >=0$.
In order to solve for the operator $f^a$, we first note that
\begin{equation}
f^a=-{W^a}_0-g \; f^{abc}\;{ (\vec \bigtriangledown ^2)}^{-1}
({W^b}_i \; \partial _i {W^c}_0).
\end{equation}
Proceeding as earlier \cite{pramana} with a mean
field type of approximation we obtain,
\begin{eqnarray}
\vec \bigtriangledown ^2{W^a}_0 (\vec x )
&&+ g^2 \; f^{abc}f^{cde} \;<vac'\mid  {W^b}_i(\vec x ) \partial _i
(\vec \bigtriangledown ^2)^{-1}({W^d}_j(\vec x ) \mid vac'>
\partial _j{W^e}_0(\vec x ))\nonumber\\ && ={J^a}_0(\vec x ),
\end{eqnarray}
where,
\begin{equation}
J^a_0=gf^{abc}{W^b_i}^{T}{G^c_{0i}}-g\bar \psi \gamma^0 \frac {\lambda ^a}{2}
\psi.
\end{equation}
\noindent
As noted earlier \cite{pramana,amspm}, the solution of equation (13)
does not suffer from Gribov ambiguity \cite{gribov}.
Clearly the solution for $W_0^a(\vec x)$ will depend on the
ansatz for the ground state $|vac'>$.

We shall now consider a trial state with gluon as
well as quark condensates. We thus explicitly take the ansatz
 for the above state as \cite{pramana,pot,njl,amspm}
\begin{equation}
|vac'>=U_GU_F|vac>,
\label{vacp}
\end{equation}
obtained through the unitary operators  $U_G$  and  $U_F$  on  the
perturbative
vacuum.
For the gluon sector, we have \cite{pramana}
 \begin{equation}
U_G
=\exp{({B_G}^\dagger-B_G)},
\end{equation}
 with
the gluon condensate creation operator ${B_G}^\dagger $
as given by
\cite{pramana}
\begin{equation}
{{B_G}^\dagger}={1\over 2}
\int {f(\vec k){{{a^a}_{i}(\vec k)}^\dagger}
{{{a^a}_{i}(-\vec k)}^{\dagger}}d\vec k},
\end{equation}
\noindent
where $f(\vec k)$ describes the vacuum structure with
gluon condensates.
For fermionic sector we have,
 \begin{equation}
U_F
=\exp{({B_F}^{\dagger}-B_F)},
\end{equation}
 with \cite{pot,njl,amspm}
 \begin{equation}
 {B_F}^{\dagger}=
\int \bigg[h(\vec k){{c^i}_{I}(\vec k)}^{\dagger}
(\vec \sigma \cdot \hat k)
{\tilde c}^i_{I}(-\vec k)
\bigg]\; d\vec k,
\label{bfbeta}
\end{equation}
Here $h(\vec k)$ is a trial function associated
with quark antiquark condensates.
 We shall minimise the energy density for $|vac^{'}>$
 to analyse vacuum stability.
For this purpose we first note that with the above transformation
the operators corresponding to $\mid vac'>$ are related to the
operators corresponding to $|vac>$ through the
Bogoliubov transformations
\begin{equation}
\left(
\begin{array}{c} b_i^a(\vec k)\\b_i^a(-\vec k)^\dagger
\end{array}
\right)
=\left(
\begin{array}{cc}
coshf(\vec k) & -sinhf(\vec k)\\
-sinhf(\vec k) & coshf(\vec k)
\end{array}
\right)
\left(
\begin{array}{c} a_i^a(\vec k)\\a_i^a(-\vec k)^\dagger
\end{array}
\right)
\end{equation}
for the gluon sector and
\begin{equation}
\left(
\begin{array}{c} d_{I}(\vec k)\\{\tilde d}_{I}(-\vec k)
\end{array}
\right)
=\left(
\begin{array}{cc}
cos(h(\vec k)) & -(\vec \sigma \cdot \hat k)sin(h(\vec k))
\\(\vec \sigma \cdot \hat k)
sin(h(\vec k)) & cos(h(\vec k))
\end{array}
\right)
\left(
\begin{array}{c}
c_{I}(\vec k)\\
{\tilde c}_{I}(-\vec k)
\end{array}
\right ),
\label{chi8}
\end{equation}
for the quark sector.

 Our job now is to evaluate the expectation value of
${\cal T}^{00}$
with respect to $\mid vac^{'}>$.
For evaluating the same, the following formulae will be useful.
\begin{equation}
<vac'\mid :{W^a}_{i}(\vec x){W^b}_{j}(\vec y):
\mid  vac'>=
{\delta }^{ab}
\times (2  \pi )^{-3}\int d\vec k e^{i\vec k.(\vec x-
\vec y)}\; {F_{+}(\vec  k)\over \omega (k)}\;
\Delta _{ij}(\vec k),
\label{ww}
\end{equation}
\begin{equation}
{<vac'\mid}: {^{T} G^{a}_{0i}} (\vec x)
{^{T} G^{b}_{0j}} (\vec y):{\mid vac'>}
= \delta ^{ab}\times (2 \pi )^{-3}
\int d{\vec k}e^{i{\vec k}.{(\vec x-\vec y)}}
{\Delta _{ij}(\vec k)\omega (k)}
F_{-}( k).
\label{gg}
\end{equation}
In the above
 $F_{\pm}(k)$ are given as \cite{pramana}
\begin{equation}
F_{\pm}(\vec k)  = \sinh^{2}f(k)
\pm{\sinh 2f(k)\over 2}
\label{fpm}
\end{equation}
Similarly, for the quark fields we have the parallel equations given as
\begin{mathletters}
\begin{equation}
<:\psi^{i}_\alpha(\vec x)^{\dagger}
\psi^{j}_\beta(\vec y):>_{vac'}=
(2\pi)^{-3}\delta^{ij}\int
\Big ( \Lambda _-(\vec k)\Big )_{\beta\alpha}
e^{-i\vec k .(\vec x-\vec y)}d\vec k,
\label{jpj}
\end{equation}
\begin{equation}
<:\psi^{i}_\alpha(\vec x)
\psi^{j}_\beta(\vec y)^{\dagger}:>_{vac'}=
(2\pi)^{-3}\delta^{ij}\int
\Big ( \Lambda _+(\vec k)\Big )_{\alpha \beta}
e^{i\vec k .(\vec x-\vec y)}d\vec k,
\label{jjp}
\end{equation}
\end{mathletters}
\noindent where
\begin{equation}
\Lambda_{\pm}(\vec  k)=\pm\frac{1}{2}\big
(\gamma ^0 \sin 2 h(\vec k)-2
(\vec\alpha \cdot \hat k)~\sin ^2 h(\vec k)\big ).
\end{equation}
These relations will be used to evaluate the energy expectation
value which is carried out in the next section.
\section{Extremisation of energy functional and results}
We shall here proceed to evaluate the expectation value of the
Hamiltonian of equation (11). However for that we note that we have to
know the  auxilary field contribution $(1/2) \partial_i f^a\partial_i f^a$
in equation (11) or equivalently the contribution arising out of the
time like and the longitudinal components of the gauge field. As stated
earlier  we shall take a mean field type of
approximation \cite{pramana} to solve for $W_0^a$
field as in equation (13). Then the solution for $W_0^a$ field
in equation (13) is given as
\cite{pramana}
\begin{equation}
{{\tilde W}^a}_0(\vec k) = \frac{{J^a}_0(\vec k)}{k^2+\phi(\vec k)}
\end{equation}
where ${\tilde W}_0^a(\vec k)$ and ${\tilde J}_0^a(\vec k)$ are
Fourier transforms of $ W_0^a(\vec x)$ and $ J_0^a(\vec x)$ and,
$\phi(\vec k)$ given through
\begin{equation}
\phi  (k)=  {3g^2\over  {8 \pi  ^2}}
\int {{dk'}\over {\omega  (k')}}\;F_{+}(k')
\biggl (  k^2+{k'}^2-{(k^2-{k'}^2)^2
\over{2kk'}}\log \Big | {{k+k'}\over{k-k'}}
\Big |  \biggr ).
\end{equation}

Using equations (\ref{t00}), (\ref{ww}), (\ref{gg}) and (25),
we then obtain
the expectation value of ${\cal T}^{00}$ with respect to
$\mid vac^{'}>$ as
\begin{eqnarray}
\epsilon_{0} &
\equiv & <vac^{'}\mid
:{\cal T}^{00}:\mid vac^{'}> \nonumber \\
& = & C_{F}
+C_{1}+C_{2}+{C_{3}}^{2}+C_{4},
\label{enrgd}
\end{eqnarray}
\noindent where \cite{pramana,pot,njl}
\begin{mathletters}
\begin{eqnarray}
C_F&=&<:\bar \psi (-i\gamma^i\partial _i
)\psi:>_{vac'}\nonumber\\ &=&
\frac {12N_f}{(2\pi)^3}\int d\vec k |\vec k| \sin ^2 h(\vec k),
\end{eqnarray}
\begin{eqnarray}
C_{1} & = & <:{1\over 2}
{^T}{G^a}_{0i}{^T}{G^a}_{0i}:>_{vac^{'}}\nonumber \\
& = & {4\over {\pi^2}}\int \omega(k)k^{2} F_{-}(k )\;dk,
\end{eqnarray}
\begin{eqnarray}
C_{2} & = & <:{1\over 2}
{W^a}_{i}{(-\vec \bigtriangledown^2)}{W^a}_{i}:>_{vac'}\nonumber
\\ & = & {4\over {\pi^2}}\int {{k^{4}}\over \omega(k)}
\;F_{+}(k )\;dk
\end{eqnarray}
\begin{eqnarray}
  {C_{3}}^{2} & = & <:{1\over 4}g^{2}f^{abc}f^{aef}
{W^b}_{i}{W^c}_{j}{W^e}_{i}{W^f}_{j}:>_{vac'}\nonumber \\
& = & \left({{2g}\over {\pi^2}}\int {{k^{2}}\over
{\omega(k)}}\;F_{+}(k)\;dk\right)^2 ,
\end{eqnarray}
\noindent and
\begin{eqnarray}
C_{4} & = & <:{1\over 2}
(\partial_{i}f^{a})(\partial_{i}f^{a}):>_{vac{'}},\nonumber\\
& = & 4\times (2 \pi )^{-6}\int  d \vec  k {G_1(\vec k)
+G_2(\vec k)\over {k^2+\phi (k)}}.
\end{eqnarray}
\end{mathletters}
In the above,
\begin{mathletters}
\begin{eqnarray}
G_1(\vec k) & = & 3 g^2 \int d  \vec q
F_{+}({\mid} \vec q\mid)\; F_{-}({\mid} \vec k +\vec q {\mid}) \;
{\omega ({\mid  \vec  k +\vec q \mid})\over \omega  ({\mid \vec  q\mid} )}
\nonumber \\ & \times & \Bigl  (1+{{(q^2 +\vec k.\vec q)^2
}\over{q^2(\vec k+\vec q)^2}}\Bigr ),
\end{eqnarray}
and, the contribution from quarks,
\begin{eqnarray}
G_2(\vec k)&=&-\frac{N_f}{2} g^2\int d \vec q
\big [ \sin 2 h(\vec q) \sin 2h(\vec k+\vec q)\nonumber \\
&+&4~\frac {\vec q \cdot (\vec k +\vec q)}{|\vec q||\vec k +\vec q|}
\sin ^2 h (\vec q) \sin ^2 h(\vec k+\vec q)\big ]
\end{eqnarray}
\end{mathletters}
and $\phi(\vec k)$ as given earlier in equation (28).
As may be noted here, the contributions from the quark
condensates to the energy density comes through auxiliary
equation through $W_0^a$ as well as from the quark kinetic term.

We shall now minimise the energy functional $\epsilon _0$ of equation
(29). For the same we shall take $\omega (\vec k)$ to be
of the free field form with an effective mass parameter
 for the gluon fields given as
\begin{equation}
\omega (\vec k)=\sqrt {k^2+m_G^2}.
\end{equation}
\noindent Here the gluon mass parameter $m_G$ given through
the self consistency condition
 \cite {pramana}
\begin{eqnarray}
  m_G^2 =
{{2g^2}\over {\pi^2}}\int {{k^{2}}\over
{\omega(k)}}\;F_{+}(k)\;dk\label{mg0}
\end{eqnarray}
arising from the sum of the single contractions of the quartic
gluon field interaction terms of ${\cal T}^{00}$ in equation (11)
\cite{pramana,biro}.

The condensate functions $f(\vec k)$ and $h(\vec k)$ are to be
determined such that the energy density $\epsilon _0$ in (29)
is a minimum. In simple cases it could be possible to solve for
the condensate functions through functional differentiation
\cite{njl,hmgrnv}. In the present case however the dependence of the
energy density on the condensate functions is highly nonlinear
and it is not possible to solve for the same  through functional
differentiation and equating it to zero. We therefore adopt here
the alternative procedure of taking a reasonably simple ansatz
for the condensate functions by parameterizing the same. We
parameterize the gluon condensates as, with $k=|\vec k|$,
\begin{equation}
\sinh f(\vec k)=Ae^{-Bk^2/2},
\end{equation}
which corresponds to taking a gaussian distribution for the
perturbative gluons in the nonperturbative vacuum
\cite{pramana}.
Similarly for the function $h(\vec k)$ describing the quark antiquark
condensates we take the ansatz
\begin{equation}
\tan 2h(\vec k)=\frac{A'}{(e^{R^2 k^2} - 1 )^{1/2}}.
\end{equation}
\noindent
The above is a generalisation of the ansatz
of ref. \cite{amspm} which corresponds to $A~'=1$
and vanishes when $A~'=0$.
It will be determined through energy minimisation.
We note that the for free massive fermions
$\tan 2h(\vec k)=m/|\vec k|$,
so that the above ansatz corresponds to a
momentum dependent mass given as
\begin{equation}
m(k)=\frac{k A'}{(e^{R ^2 k^2} - 1 )^{1/2}}.
\end{equation}
We may add here that such a definition of quark mass is the same as that
obtained from the pole of the fermion propogator in a condensate vacuum
\cite{pot,mac}.
In the limit of zero momentum we then have the dynamically
generated mass for the quarks given as
\begin{equation}
m_{q}=\frac{A'}{R}.
\label{mq}
\end{equation}
The relationship
between the decay constant of pion and the quark condensate function
has been discussed in ref. \cite{amspm} and is given as
\begin{equation}
\frac{N_\pi}{(2\pi)^{3/2}}\int \sin^2 2h(k) d\vec k=\frac{c_\pi(m_\pi)^{1/2}}
{\sqrt 6}
\end{equation}
where $N_\pi \sin 2h(k)$ is the wave function for the pion with
\begin{equation}
N_\pi^{-2}=\int \sin^2 2 h(k) d\vec k.
\label{npi}
\end{equation}
With the ansatz of equation (35) we then have
\begin{equation}
R= \left(  \frac{\sqrt 3}{\pi c_\pi \sqrt{m_\pi}}
\right)^{2/3}.
\left[ \int \frac{{A'}^2 x^2 dx }{e^{x^2} +1-{A'}^2 } \right]^{1/3} .
\end{equation}
The above equation determines $R$ as a function of $A'$ when $c_\pi$ is
known. This will be used when we extremise energy along with a parallel
constraint for SVZ parameter as in equation (48).

With the above ansatzes for the condensate functions the energy
density $\epsilon_{0}$ may now be written in terms of the dimensionless
quantities  $x=Rk$ and $b=B/R^2$ as
\begin{eqnarray}
\epsilon_{0}(A)
& = & {1\over R^4}(I_F+I_{1}(A,b)+I_{2}(A,b)+{I_{3}(A,b)}^{2}+
I_{4}(A,b)) \nonumber \\
& \equiv & {1\over R^4}F(A,b),
\end{eqnarray}
where \cite{pramana,pot,njl}
\begin{mathletters}
\begin{equation}
I_F=\frac{3N_f}{\pi^2}\int x^3 dx
\bigg[1-\bigg(1- \frac{A'^2}{e^{x^2}-1+A'^2}\bigg)^{1/2}\bigg],
\end{equation}
\begin{equation}
I_1(A,b)=\frac{4}{\pi^2}\int \omega(x,b)x^2 dx
\bigg(
A^2e^{-bx^2}-Ae^{-bx^2/2}\big(1+A^2e^{-bx^2}\big)^{1/2}
\bigg),
\end{equation}
\begin{equation}
I_2(A,b)=\frac{4}{\pi^2}\int\frac{x^4 dx }{\omega(x,b)}
\bigg(
A^2e^{-bx^2}+Ae^{-bx^2/2}\big(1+A^2e^{-bx^2}\big)^{1/2}
\bigg)
\end{equation}
\begin{equation}
I_3(A,b)=\frac{2g}{\pi^2}\int\frac{x^2 dx }{\omega(x,b)}
\bigg(
A^2e^{-bx^2}+Ae^{-bx^2/2}\big(1+A^2e^{-bx^2}\big)^{1/2}
\bigg)
\end{equation}
\noindent  and
\begin{equation}
I_{4}(A,b) =4\times {(2 \pi )^{-6}}\int d{\vec x}
{{G_1(\vec x)+G_2(\vec x)}\over {x^2+\phi (x)}}.
\label{i4}
\end{equation}
\end{mathletters}
with
\begin{eqnarray}
G_1(\vec x) & = & 3g^2\int d\vec x^{'}
\bigg(
A^2e^{-b{x'}^2}+Ae^{-b{x'}^2/2}\big(1+A^2e^{-b{x'}^2}\big)^{1/2}
\bigg)\nonumber \\ &\times &
\bigg(
A^2e^{-b(\vec x+{\vec x'})^2}+Ae^{-b(\vec x+{\vec x'})^2/2}
\big(1+A^2e^{-b(\vec x+{\vec x'})^2}\big)^{1/2}
\bigg)\nonumber\\ &\times &
\frac{\omega(|\vec x+{\vec x'}|)}{\omega(x')}\times
\bigg (1+\frac{{({\vec x~'}^2+\vec x \cdot \vec x~')}^2}{{x~'}^2
{(\vec x+\vec x~')}^2}\bigg),
\end{eqnarray}
\begin{eqnarray}
G_2(\vec x) & = & -\frac{N_f}{2}g^2\int d\vec x~'
\bigg[ \frac{A'^4}{(e^{x'^2}-1+A'^2)
(e^{(\vec x+\vec x')^2}-1+A'^2)}+
\frac{\vec x~' \cdot (\vec x +\vec x~')}{|\vec x~'| |\vec x+\vec x~'|}
\nonumber \\
& \times &
\bigg (1-{\bigg(1-\frac{A'^2}{(e^{x'^2}-1+A'^2)}\bigg
)}^{1/2}\bigg)
\bigg (1-\bigg(1-\frac{A'^2}{(e^{(\vec x+\vec x')^2}-1+A'^2)}
\bigg)^{1/2}\bigg) \bigg],
\end{eqnarray}
and
\begin{eqnarray}
\phi (\vec x) & = & \frac{3g^2}{8\pi^2}
\int \frac{d x'}{\omega({\vec x'})} \bigg(
A^2e^{-b{x'}^2}+Ae^{-b{x'}^2/2}\big(1+A^2e^{-b{x'}^2}\big)^{1/2}
\bigg)\nonumber \\ &\times &
\bigg(
x^2+{x'}^2-\frac{(x^2-{x'}^2)^2}{2xx'}log\Big|\frac{x+x'}
{x-x'}\Big|\bigg).
\end{eqnarray}
In the above, $\omega(x,b)=(x^2+\mu'^2)^{1/2}$,
with $\mu'=m_G R$ being the gluon mass.
The self consistency condition \cite{pramana} for gluon mass $m_G$ in equation
(33) yields
\begin{equation}
\mu^2=\frac{2g^2}{\pi^2}\int \frac{x^2 dx}{(x^2+\mu^2)^{1/2}}
\bigg(
A^2e^{-x^2}+Ae^{-x^2/2}\big(1+A^2e^{-x^2}\big)^{1/2}
\bigg),
\end{equation}
where $\mu=m_G \sqrt B$. Clearly
$\mu'$ and $\mu$ are related as $\mu' = \mu/\sqrt b $.
For given values of $A$ and $A'$, we first determine $R\equiv R(A~')$
from equation (39),
and then the parameter $b$ from the SVZ parameter for gluon condensates
through the equation
\begin{equation}
\frac{g^2}{4\pi^2}<:G_{\mu\nu}^a G^{a\mu\nu}:>_{vac'}
=0.012\; GeV^4,
\end{equation}
which  explicitly gives  \cite{pramana}
\begin{equation}
\frac{1}{R(A~')^4} \times\frac{g^2}{\pi^2}\big(-I_1(A,b)+I_2(A,b)+I_3(A,b)^2
-I_4(A,b)\big)=0.012\; GeV^4,
\end{equation}
so that $b$ is a function of $A$ and $A~'$.

We now minimise the energy density $\epsilon_0$ with respect
to the parameters $A$ and $A'$ with a self consistent determination
for the gluon mass as in equation (47) for different values of the
QCD coupling constant $\alpha_s$.
We plot in curve 1 of Fig.1  $\epsilon_0$ as a function of $\alpha_s$.
We note that for $\alpha_s \leq 0.62=\alpha_c$ the condensate functions
vanish, so that for the present ansatz, perturbative vacuum is stable, whereas,
for $\alpha_s >\alpha_c$, we have a transition to a nonperturbative vacuum.
The energy density becomes more negative with
increase of coupling and becomes about $-$55 MeV/fm$^3$ when $\alpha_s=1.4$.
In curve 2 of the same figure we have plotted gluon condensate parameter
$A_{min}$ for different couplings. In contrast to ref.\cite{pramana},
near $\alpha_s=\alpha_c$ this parameter has a discontinuity.
In curve 3 we have plotted ${\sqrt B}$
in units of fermi as a function of coupling.
The length scale here is of
the order of a fermi which appears to be reasonable in the context
of confinement. In curve 4, we have plotted the gluon mass parameter $m_G$
in MeV as a function of coupling.
This quantity appears to be constant and is around 300 MeV as earlier
\cite{pramana}, and is similar to the result of Cornwall \cite{corn}
where we adjust for the QCD parameter $\Lambda$ or,
 Monte Carlo simulations of lattice calculations\cite{mandula}.
 For such gluons QGP signals are considered for
 relativistic heavy ion collisions \cite{biro}.

In Fig. 2 we have plotted different characteristics of the quark
condensate functions. In curves 1 and 2 we have plotted the condensate
parameters $A'_{min}$ and $R$ respectively. They seem to increase as
we approach $\alpha_c$ from above.
We have plotted  in curve 3 the quark condensate value given as
\begin{equation}
<-\bar \psi\psi>^{1/3}=\Big[\frac{12}{(2\pi)^3}\int d\vec k \sin 2h(\vec k)
\Big]^{1/3}.
\end{equation}
The condensation effect increases
with coupling as expected. In curve 4 we have plotted the
dynamically generated quark mass $m_q$ of equation (\ref{mq}).
We may in fact identify the same as the parallel of
constituent mass of quarks
as obtained from chiral symmetry breaking.

We may remark that for $\alpha_s<\alpha_c$, any nonzero trial functions
make $\epsilon_0$ positive, which shows that for the present ansatz
the perturbative vacuum is stable.

\section{Discussions}

We have considered here destabilisation of perurbative vacuum in QCD
as a function of coupling constant through an explicit construction. We have
taken an ansatz for nonperturbative vacuum with trial functions
both for quark and gluon condensates. With the QCD Hamiltonian
for quarks and gluons satisfying equal time algebra, we then do
a minimisation of the energy density with the constraints that the
pion decay constant and the SVZ parameters are correctly reproduced.
Minimisation of the energy density over remaining parameters
yields that vacuum with condensates is the preferred configuration when 
$\alpha_s$ is greater than 0.62. In order to relate the vacuum structure
of chiral symmetry breaking to $c_\pi$, we exploit the results of
Ref.\cite{pot,njl,amspm}.

It shall be amusing to consider the present results in the context of
Ref.\cite{amspm} where {\it no} extremisation was done. We find that for
$\alpha_s=1.28$, $A'_{min}\simeq 1$, and that the pion charge radius
gets correctly reproduced. In fact, with $A'=1$, we have \cite{amspm}
\begin{equation}
R_{ch}^2=R_1^2+R_2^2,
\end{equation}
where
\begin{equation}
R_1^2=\frac{N_\pi^2}{4}\int|\vec \bigtriangledown\sin 2h(k)|^2 d\vec k
\end{equation}
and
\begin{equation}
R_2^2=\frac{N_\pi^2}{16}\int\sin^2 2h(k)
\Big[k^2R^4 \tan^2 2 h(k)+ \frac{4(1-\sin 2h(k)}{k^2}\Big]d\vec k
\end{equation}
In the above $N_\pi$ is given in equation (\ref{npi}).
With $R\simeq 0.96$ fm, we then obtain that $R_{ch}\simeq 0.65$ fm,
which agrees with the experimental value of 0.66 fm \cite{amendolia}
The new feature here is that we are able to relate the QCD coupling constant
with low energy hadron properties.

We may also note that the averge number of perturbative
gluons or quarks in such a condensate vacuum is given as
\begin{equation}
N_G=<vac'|{a_i^a(\vec z)}^\dagger a_i^a(\vec z)|vac'>
= 2\times 8\times\frac{1}{2\pi^3}\int \sinh ^2f(\vec k) d\vec k.
\end{equation}
and,
\begin{eqnarray}
N_q&=&\frac{12}{(2\pi)^3}\int \sin^2 h(\vec k)d\vec k\nonumber\\
&=&\frac{1}{R^3}\frac{3}{\pi^2}\int x^2dx\bigg [ 1-
\bigg ( 1-\frac{A'^2}{e^{x^2}-1+A'^2}\bigg )^{1/2}\bigg ].
\end{eqnarray}
We then obtain that for example when $\alpha_s=1.28$
corresponding to correct charge radius of pion,
$N_G={2A^2}/{(\pi B)^{3/2}}\simeq 0.233$/fm$^3$ and that
$N_q\simeq 0.085$/fm$^3$.
We also note that for $\alpha_s=1.28$, the ``bag pressure''
is given as $(-\epsilon_0)^{1/4}=140$ MeV, which appears
to be in general agreement with the phenomenological value
of this parameter.

We thus find that a constrained energy minimisation of QCD
Hamiltonian can lead to both gluon and quark condensates
with appropriate chiral symmetry breaking.
Even we can have simultaneously both the charge radius
and pion decay constant, features of chiral symmetry breaking,
getting related to the QCD coupling constant.
The method in fact seems to open up
possibilities of linking up the (mysterious) vacuum structure of QCD to
(common place) hadronic properties in an unexpected manner.

We have focussed our attention on the structure problem.
It will be desirable to include the dynamical effects
in the light quark sector. Also, chiral symmetry breaking is
important in the strange quark sector, and calculations adding
condensates of the same can be carried out. Besides
technical difficulty, this has the additional feature
of Lagrangian mass being of the same order as dynamical mass
and may need nontrivial extension of the technology.
Work in this direction are in progress.
\acknowledgements
The authors are thankful to J.C. Parikh, S.B. Khadkikar and N. Barik
for many useful discussions.
SPM would like to thank Department of
Science and Technology, Government of India for
research grant no SP/S2/K-45/89 for financial assistance.

\def \sur {E.V. Shuryak, Rev. Mod. Phys. 65, 1 (1993);
E.V. Shuryak, {\it The QCD vacuum,
hadrons and the superdense matter}, (World Scientific, Singapore).}
\def \qcd {G. K. Savvidy, Phys. Lett. 71B, 133 (1977);
S. G. Matinyan and G. K. Savvidy, Nucl. Phys. B134, 539 (1978); N. K. Nielsen
and P. Olesen, Nucl.  Phys. B144, 376 (1978).}
\def \hans {T. H. Hansson, K. Johnson,
C. Peterson, Phys. Rev. D26, 2069 (1982).}
\def \corn{J. M. Cornwall, Phys. Rev. D26, 1453, (1982).}
\def \njl {H.Mishra and S.P. Misra, Phys Rev. D48, 5376 (1993).}

\def \svz {M.A. Shifman, A.I. Vainshtein and V.I. Zakharov,
Nucl. Phys. B147, 385, 448 and 519 (1979);
R.A. Bertlmann, Acta Physica Austriaca 53, 305 (1981).}

\def \nambu{ Y. Nambu, Phys. Rev. Lett. 4, 380 (1960);
Y. Nambu and G. Jona-Lasinio, Phys. Rev. 122, 345 (1961); ibid,
124, 246 (1961);
J.R. Finger and J.E. Mandula, Nucl. Phys. B199, 168 (1982);
A. Amer, A. Le Yaouanc, L. Oliver, O. Pene and
J.C. Raynal, Phys. Rev. Lett. 50, 87 (1983);
ibid, Phys. Rev. D28, 1530 (1983); S.L. Adler and A.C. Davis,
Nucl. Phys. B244, 469 (1984); R. Alkofer and P. A. Amundsen,
Nucl. Phys.B306, 305 (1988); A.C. Davis and A.M. Matheson,
Nucl. Phys. B246, 203 (1984);
S. Schramm and W. Greiner, Int. Jour. Mod. Phys. E1, 73
(1992).}
\def \mac{M. G. Mitchard, A. C. Davis
and A. J. Macfarlane, Nucl. Phys. B325, 470 (1989).}

\def \pot{A. Mishra, H. Mishra and S. P. Misra, Z. Phys. C57, 241 (1993).}

\def \lat{K.G. Wilson, Phys. Rev. D10, 2445 (1974); J.B. Kogut, Rev. Mod.
Phys. 51, 659 (1979); ibid 55, 775 (1983); M. Cruetz, Phys. Rev. Lett.
45, 313 (1980); ibid Phys. Rev. D21, 2308 (1980); T. Celik, J. Engels and
H. Satz, Phys. Lett. B129, 323 (1983); H. Satz, in Proceedings of Large Hadron
Collider Workshop, Vol.I, Ed. G. Jarlskog and D. Rein, CERN 90-10, ECFA 90-133
(1990).}

\def \biroo {T. Biro, Ann. Phys. 191, 1 (1989), Phys. Lett B228, 16 (1989);
Phys. Lett. B245, 142 (1990).}
\def \joglekar {J. Collins, A. Duncan and S. Joglekar, Phys. Rev.
D16, 438 (1977); N.K. Nielsen, Nucl. Phys. B120, 212 (1977);
J. Schechter, Phys. Rev. D21, 3393 (1981); A.A. Migdal
and M.A. Shifman, Phys. Lett. 114B, 445 (1982).}

\def \star {H. Mishra,
S.P. Misra, P.K. Panda and B.K. Parida, Int. J. Mod. Phys.
E2, 547 (1993); H. Mishra, A. Mishra, S.P. Misra and P.K. Panda,
preprint hepph/9401344.}

\def \pramana { A. Mishra, H. Mishra, S.P. Misra and S.N. Nayak,
Pramana (J. of Phys.) 37, 59 (1991);
A. Mishra, H. Mishra, S.P. Misra
and S.N. Nayak, Z. Phys. C57, 233 (1993);
A. Mishra, H. Mishra and S.P. Misra, Z. Phys. C59, 159 (1993).}

\def \tfd {H. Umezawa, H. Matsumoto and M. Tachiki,{\it
Thermofield Dynamics and Condensed States} (North Holland,
Amsterdam, 1982).}

\def \higgs {A. Mishra, H. Mishra, S.P. Misra and S.N. Nayak, Phys. Rev.
D44, 110 (1991).}
\def \nmtr {H. Mishra,
S.P. Misra, P.K. Panda and B.K. Parida, Int. J. Mod. Phys. E1, 405 (1992);
P.K. Panda, R. Sahu and S. P. Misra, Phys. Rev. C45, 2079 (1993); H. Mishra,
S.P. Misra, P.K. Panda and B.K. Parida, Int. J. Mod. Phys.
E2, 547 (1993); H. Mishra, A. Mishra, S.P. Misra and P.K. Panda,
preprint hepph/9401344.}

\def \shut{D. Schutte, Phys. Rev. D31, 810 (1985).}

\def \schwinger{  J. Schwinger, Phys. Rev. 125, 1043 (1962); ibid,
127, 324 (1962); E. S. Abers and B. W. Lee, Phys. Rep. 9C, 1 (1973). }

\def \spm { S. P. Misra, Phys. Rev. D18, 1661, 1673 (1978);
A. Le Youanc et al Phys. Rev. Lett. 54, 506
(1985).}

\def \hmgrnv { H. Mishra, S.P. Misra and A. Mishra,
Int. J. Mod. Phys. A3, 2331 (1988);
S.P. Misra, Phys. Rev. D35, 2607 (1987).}
\def \amspm {A. Mishra and S.P. Misra, Z. Phys. C58, 325 (1993).}

\def \fetter {A. L. Fetter and J. D. Walecka, {\it Quantum Theory of
Many Particle Systems}, McGraw Hill Book Company, 1971.}

\def \kapusta {J.I. Kapusta, {\em Finite Temperature
Field Theory }(Cambridge University Press 1989);
B. Freedman and L. McLerran, Phys. Rev. D17, 1109
(1978); B. D. Serot and H. Uechi, Ann. Phys. 179, 272 (1987).}

\def \biro {Tamas S Biro, Int. J. Mod. Phys. E1, 39 (1992).}

\def \satz {H. Satz, Nucl. Phys. A498, 495c (1989).}

\def \quenched {D. Barkai, K.J.M. Moriarty and C. Rebbi,
Phys. Rev. D30, 1293 (1984); A.D. Kennedy, J. Kuti, S.Meyer and B.J.
Pendelton,
Phys. Rev. Lett. 54, 87 (1985).}
\def \unquenched {S. Gottlieb et al, Phys. Rev. D38, 2245 (1988).}
\def \gavai {R.V. Gavai in Proceedings of Workshop on Role of Quark
Matter In Physics and Astrophysics, edited by R.S. Bhalerao and R.V. Gavai,
Bombay 1992.}
\def \witten{ E. Witten, Phys. Rev. D30, 272 (1984);
E. Farhi and R. L. Jaffe, Phys. Rev. D30, 2379 (1984).}
\def \glendn {N.K. Glendenning, Talk presented at the Workshop on
Strange Quark Matter in Physics and Astrophysics, Aarhus, Denmark, 1991, to
appear in the proceedings; ibid preprint LBL-30878.}
\def \ellis{J. Ellis, J.I. Kapusta and K.A. Olive, Nucl. Phys. B348,
345 (1991); ibid, Phys. Lett. B273, 123 (1991); G. E. Brown and Mannque Rho,
Phys. Rev. Lett. 66, 2720 (1991).}
\def \astro {T. \O verg\.{a}rd and E. \O verg\.{a}rd,
Astro. Astrophys. 243, 412 (1991); ibid, Class.
Quantum Grav. 8, L49, 1991; B.Datta, P. K. Sahu, J. D. Anand
and A. Goyal, Phys. Lett. B283, 313 (1992).}
\def\rosen {A. Rosenhauer et al  Nucl. Phys. A540, 630 (1992).}
\def\weinberg {S. Weinberg, {\it Gravitation and cosmology}, (Wiley,
New York, 1972);
F. Weber and M.K. Weigel, Nucl. Phys. {\bf A493} (1989) 549.}
\def\shapiro {S. L. Shapiro and S. A. Teulosky, {\it Black holes,
white dwarfs and neutron stars} (Wiely, New York, 1983).}
\def\brecher {K. Brecher, Astro. Phys. J. {\bf 215} (1977) L17.}
\def\mandula{J.E. Mandula, M. Ogilvie, Phys. Lett. B185, 127 (1987).}
\def\gribov{V. N. Gribov, Nucl. Phys. B139, 1 (1978).}
\def\amendolia{S.R. Amendolia {\it et al}, Nucl. Phys. B277, 168 (1986).}

 \newpage
\centerline{\bf Figure Captions}
\bigskip
\noindent {\bf Fig.1:} In curves 1,2,3 and 4, energy density
$\epsilon_0$ (in units of 100 MeV/fm$^3$), $A_{min}$ ,
$\sqrt B$ (in units of fm) and
effective gluon mass $m_G$ (in units of 200 MeV),
are plotted respectively as functions of strong coupling constant $\alpha_s$.
\hfil\break

\medskip

\noindent {\bf Fig.2:} In curves 1,2,3 and 4, $A'_{min}$,$R$ (in units of fm),
 the quark condensate $(-<\bar\psi\psi>)^{1/3}$ (in units of 100 MeV) and
dynamically generated quark mass $m_q$ (in units of 200 MeV),
are plotted
respectively as functions of the coupling constant.
$\alpha_s$.
\end{document}